\documentclass[aps, pra, 10pt, superscriptaddress, nobibnotes, reprint]{revtex4-2}

\usepackage[T1]{fontenc}
\usepackage[utf8]{inputenc}
\usepackage[english]{babel}
\usepackage[protrusion=false]{microtype}

\usepackage{amsmath}
\usepackage{amssymb, amsfonts, mathrsfs}
\usepackage{array}

\usepackage{graphicx}
\usepackage{booktabs}
\usepackage{float}

\usepackage{physics}

\usepackage{xcolor}
\usepackage{hyperref}

\definecolor{darkblue}{rgb}{0,0,0.4}
\definecolor{lightblue}{rgb}{0,0,0.8}

\hypersetup{
    colorlinks=true,
    linkcolor=lightblue,
    citecolor=lightblue,
    urlcolor=lightblue
}

\usepackage[all]{hypcap}

\begin{document}

\title{A Degenerate Singlet–Triplet Qubit with All-Electrical Orthogonal Control }

\author{P. X. Nguyen}
\thanks{These authors contributed equally to this work and are listed in alphabetical order.}\affiliation{Department of Physics, Harvard University, Cambridge, MA, USA}

\author{K. Tsoukalas}
\thanks{These authors contributed equally to this work and are listed in alphabetical order.}\affiliation{Department of Physics, Harvard University, Cambridge, MA, USA}

\author{J. H. Ungerer}
\thanks{These authors contributed equally to this work and are listed in alphabetical order.}
\affiliation{Department of Physics, Harvard University, Cambridge, MA, USA}

\author{J. Santen}
\affiliation{Department of Physics, Harvard University, Cambridge, MA, USA}

\author{V. John}
\affiliation{QuTech and Kavli Institute of Nanoscience, Delft University of Technology, 2600 GA Delft, Netherlands}

\author{S. D. Oosterhout}
\affiliation{QuTech and Netherlands Organisation for Applied Scientific Research (TNO), 2628 CK Delft, Netherlands}

\author{L. Stehouwer}
\affiliation{QuTech and Kavli Institute of Nanoscience, Delft University of Technology, 2600 GA Delft, Netherlands}

\author{S. Bosco}
\affiliation{QuTech and Kavli Institute of Nanoscience, Delft University of Technology, 2600 GA Delft, Netherlands}

\author{G. Scappucci}
\affiliation{QuTech and Kavli Institute of Nanoscience, Delft University of Technology, 2600 GA Delft, Netherlands}

\author{M. Veldhorst}
\affiliation{QuTech and Kavli Institute of Nanoscience, Delft University of Technology, 2600 GA Delft, Netherlands}

\author{A. Yacoby}
\affiliation{Department of Physics, Harvard University, Cambridge, MA, USA}

\date{\today}

\begin{abstract}
Singlet–triplet qubits offer an attractive encoding for semiconductor quantum computing, combining ancilla-free readout, reduced sensitivity to common-mode noise, and baseband voltage control. However, the Zeeman energy difference $\Delta E_\mathrm{Z}$ is typically fixed by local magnetic field gradients or $g$-factor inhomogeneities, leaving the exchange interaction $J$ as the only dynamically tunable parameter. This always-on $\Delta E_\mathrm{Z}$ precludes orthogonal control of the qubit's rotation axes and introduces unwanted state rotations during idling. Here we demonstrate all-electrical orthogonal control of a degenerate singlet–triplet (DST) qubit formed by two hole spins in a germanium double quantum dot. Exploiting the electrically tunable anisotropic $g$-factors of the two spins, we identify a regime where both $\Delta E_\mathrm{Z}$ and $J$ vanish, making the $\ket{S}$ and $\ket{T_0}$ states degenerate at the idle point. By applying only baseband voltage pulses, we independently control both $J$ and $\Delta E_\mathrm{Z}$, enabling fully orthogonal $Z$- and $X$-axis rotations. Randomized benchmarking yields an average physical single-qubit gate fidelity of 99.53\% for a gate duration of approximately 100 ns. Finally, we electrically tune the degenerate point across a wide range of magnetic field orientations, enabling operation in a regime of enhanced coherence time and offering a route towards multi-qubit scaling under a shared global magnetic field.

\end{abstract}

\maketitle

\section{Introduction}

As semiconductor spin qubits scale to larger processors, new qubit approaches that optimize control and minimize complexity are of central interest. Conventional spin control is typically achieved through electron spin resonance, enabling high-fidelity operations, orthogonal two-axis control and precise phase tracking between pulses. Yet, this technique faces substantial challenges for large-scale integration, including RF-induced heating~\cite{Undseth2023Hotter}, crosstalk~\cite{Undseth2023} and the overhead associated with on-chip RF electronics~\cite{Xue2021,Bartee2025}. To mitigate these limitations, several alternative qubit encodings and control schemes have been developed that exploit diabatic Hamiltonian evolution for spin manipulation. These include spin shuttling between quantum dots (QD) with non-collinear quantization axes~\cite{LossDiVincenzo1998,Wang2024}, the singlet--triplet (ST) qubit~\cite{Levy2002,Petta2005}, and the exchange-only (EO) qubit~\cite{DiVincenzo2000,Medford2013,bosco2026exchange}, all controlled exclusively through baseband voltage pulses.

However, these approaches generally suffer from non-orthogonal control axes, requiring multiple sequential rotations to realize arbitrary single-qubit operations and thereby increasing gate complexity~\cite{Wang2024,Zhang2024,Weinstein2023}. Furthermore, qubit states are typically non-degenerate, resulting in unwanted residual qubit rotations that must be tracked in a rotating frame, adding the overhead of fast clocks to follow each qubit's phase. Although theoretical proposals exist for eliminating idle spin rotations~\cite{RimbachRuss2025}, only the EO qubit has achieved this experimentally at the cost of requiring three spins in a triple quantum dot~\cite{HRL2026,Madzik2025}.

The ST qubit is encoded in the singlet $\ket{S}$ and triplet $\ket{T_0}$ states of two spins in a double QD and controlled through the exchange interaction $J$ and Zeeman energy difference $\Delta E_\mathrm{Z}$ between the two spins. $J$ defines the energy splitting between $\ket{S}$ and $\ket{T_0}$ ($Z$ gate), while $\Delta E_\mathrm{Z}$ performs state rotations in the $\{\ket{S}$,$\ket{T_0}\}$ subspace ($X$ gate). This is a particularly attractive encoding, using only two QDs with built-in spin-to-charge conversion for readout without ancilla spins while also being insensitive to common-mode magnetic and charge noise~\cite{Tomic2025,Levy2002,Petta2005,Jirovec2021,liles2024singlet}.

A key limitation of the ST qubit is the restricted control over $\Delta E_\mathrm{Z}$, which is typically fixed during an experiment by local magnetic field gradients or $g$-factor inhomogeneities, leaving $J$ as the only dynamically tunable parameter. Existing approaches to adjusting $\Delta E_\mathrm{Z}$, such as micromagnet-induced gradients, nuclear-spin pumping, or changes to the global magnetic field are slow and largely incompatible with scaling~\cite{Foletti2009,Wu2014,Berritta2024,Jirovec2021,liles2024singlet}. Moreover, because $\Delta E_\mathrm{Z}$ is not dynamically tunable, increasing its value to achieve faster $X$ gates necessitates a commensurate increase in $J$ to preserve two-axis control ($J \gg \Delta E_\mathrm{Z}$), rendering the qubit more susceptible to charge noise.

In Ge/SiGe hole spin qubits, strong spin–orbit interaction results in anisotropic $g$-tensors that can be modulated electrically through gate voltages~\cite{Martinez2022,Hendrickx2024,Bulaev2007,Scappucci2021}. Although this modulation has been widely exploited to enable all-electrical resonant control of individual spins~\cite{Hendrickx2021,Dijkema2026,Hendrickx2024}, its influence on the $\Delta E_\mathrm{Z}$ of two spins has not been thoroughly explored for singlet--triplet qubits. Independent control of both $J$ and $\Delta E_\mathrm{Z}$ could enable an operating regime in which both parameters are simultaneously tuned to zero and the qubit's two computational states $\ket{S}$ and $\ket{T_0}$ become energetically degenerate~\cite{RimbachRuss2025, Hansen2024}.

\begin{figure*}[tbp]
\centering
\includegraphics[width=\textwidth]{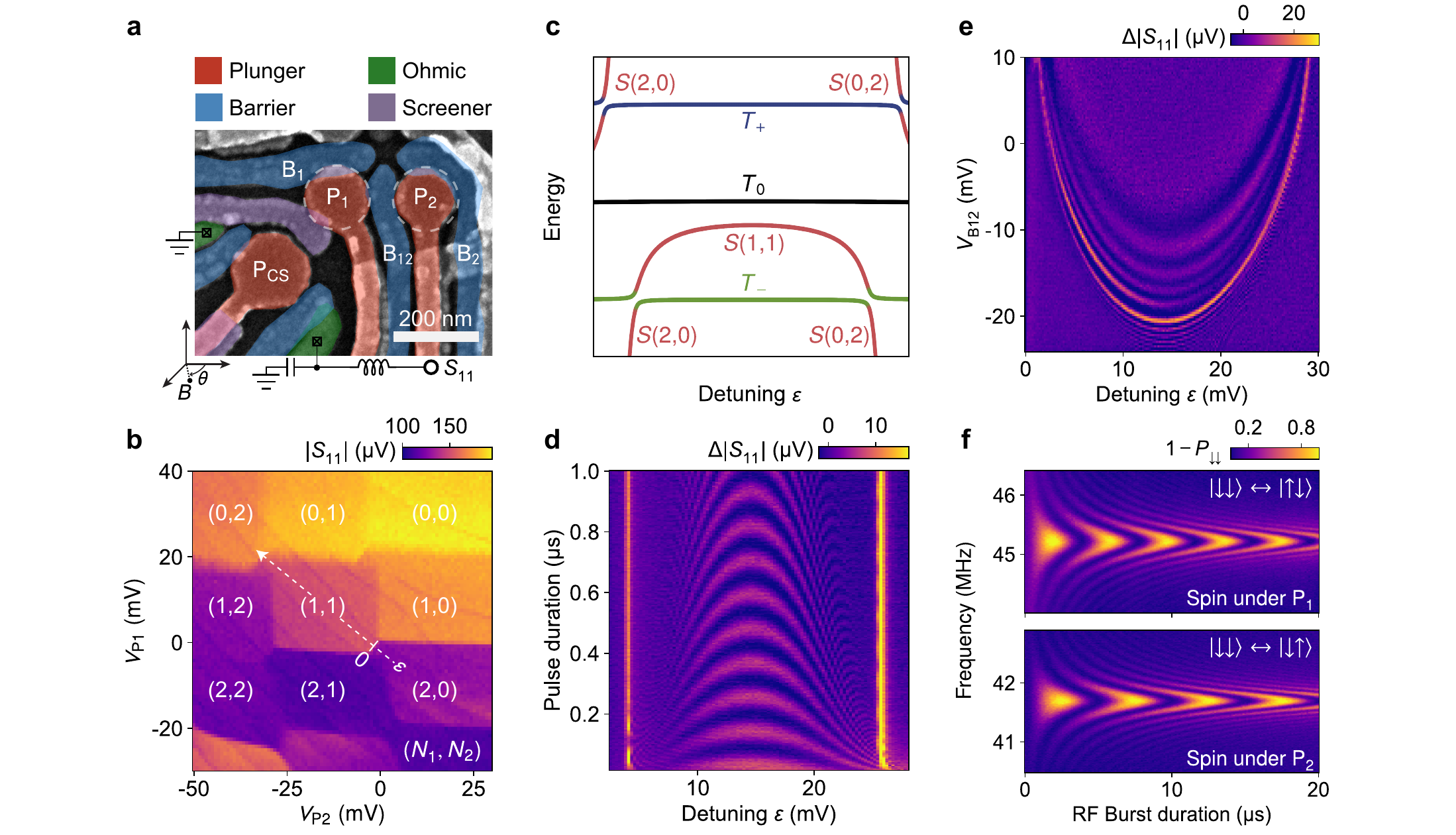}
\caption{\textbf{A two-hole spin device.}
\textbf{a}, False-colored scanning-electron microscopy image of a nominally identical device design reported in Ref.~\cite{John2025}. The quantum dots are formed under $\mathrm{P}_1$ and $\mathrm{P}_2$, while a sensing dot is formed under $\mathrm{P}_\mathrm{CS}$, with one ohmic contact connected to an RF tank circuit, where reflected $|S_{11}|$ signal is measured for readout.
\textbf{b}, Charge stability diagram measured by recording $|S_{11}|$ as the virtual plunger voltages $V_\mathrm{P1}$ and $V_\mathrm{P2}$ are varied. Charge occupations are labeled $(N_1,N_2)$. The white dashed line indicates the detuning axis in the two-hole regime.
\textbf{c}, Energy spectrum of the two-spin states as a function of detuning.
\textbf{d}, $S$--$T_0$ and $S$--$T_-$ oscillations generated by square diabatic detuning pulses of variable duration, corresponding to the energy spectrum shown in \textbf{c}. $\Delta|S_{11}|$ denotes the background-subtracted $|S_{11}|$ signal.
\textbf{e}, Baseband spectroscopy revealing the position of the $S$--$T_-$ anticrossing as a function of detuning and barrier gate voltage.
\textbf{f}, Measurement of the population of $\ket{\downarrow\downarrow}$ after performing electrically driven spin resonance as a function of RF drive frequency and burst duration at $|B|= 20$\,mT.}
\label{fig1}
\end{figure*}

Here, we demonstrate a degenerate singlet--triplet (DST) qubit with fully orthogonal two-axis control using only baseband voltage pulses in a Ge/SiGe heterostructure hosting two hole spins. We achieve this by exploiting the voltage-tunable $g$-factors of the holes, thereby eliminating the need for fixed local magnetic field gradients. Operating at a magnetic field of 40\,mT, we encode the qubit in the degenerate $\ket{S}$ and $\ket{T_0}$ states, where selective state initialization and readout are achieved through distinct pulse-ramp sequences. We independently characterize how different gate-voltage combinations tune $J$ and $\Delta E_\mathrm{Z}$, and exploit these controls to implement orthogonal $Z$- and $X$-axis rotations on the Bloch sphere. Randomized benchmarking demonstrates an average physical and Clifford gate fidelity of $F_{\text{gate}} = 99.53(9)\%$ and $F_\mathrm{C} = 98.7(3)\%$ respectively, while two distinct coherence times are extracted corresponding to the two driving axes, $T^*_{2,X}=1.55(1)$\,\textmu s and  $T^*_{2,Z}=2.32(4)$\,\textmu s. Finally, we demonstrate that the degenerate point of the DST qubit can be electrically tuned across different magnetic field orientations. This allows us to identify a regime where the hyperfine noise is suppressed, leading to significant enhancement in the coherence time $T^*_{2,X}$.

\section{Results}

\begin{figure*}[tbp]
\centering
\includegraphics[width=\linewidth]{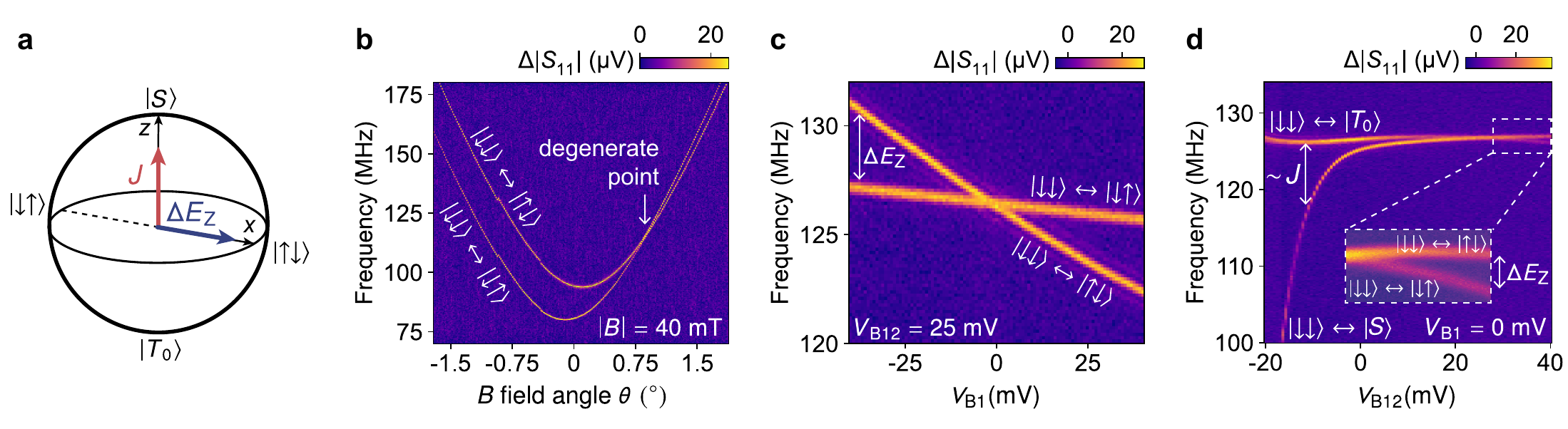}
\caption{\textbf{Control over singlet--triplet qubit parameters.}
\textbf{a}, Bloch-sphere representation of the singlet--triplet qubit. The exchange interaction $J$ and Zeeman-energy difference $\Delta E_\mathrm{Z}$ generate rotations about the $Z$- and $X$-axis, respectively.
\textbf{b}, RF spectroscopy of the spin-resonance frequencies of the two hole spins as a function of out-of-plane magnetic field angle at fixed amplitude $|B|= 40$\,mT, revealing the degenerate point at which the two resonance frequencies coincide.
\textbf{c}, RF spectroscopy of the spin-resonance frequencies as a function of the virtual gate voltage $V_\mathrm{B1}$, demonstrating electrical control of $\Delta E_\mathrm{Z}$.
\textbf{d}, RF spectroscopy of the spin-resonance frequencies as a function of the virtual gate voltage $V_\mathrm{B12}$. Negative $V_\mathrm{B12}$ increases the exchange interaction $J$, whereas positive $V_\mathrm{B12}$ generates a finite $\Delta E_\mathrm{Z}$ (inset).}
\label{fig2}
\end{figure*}

\subsection{Two-hole spin device in Ge}
The experiment is performed on parts of a ten-quantum-dot device fabricated on a Ge/SiGe heterostructure, shown in Fig.~\ref{fig1}a. Details of the heterostructure growth and device fabrication are provided in Methods~\ref{Methods:fabrication}. Here we focus on the double quantum dot (DQD) formed under the plunger electrodes (gates) $\mathrm{P}_1$ and $\mathrm{P}_2$ (see Fig.~\ref{fig1}a). Charge sensing is performed by radio-frequency reflectometry of a nearby sensor dot formed under $\mathrm{P}_\mathrm{CS}$ (Methods~\ref{Methods:reflectometry}). By varying the virtual plunger voltages $V_\mathrm{P1}$ and $V_\mathrm{P2}$ (Methods~\ref{Methods:Virtual gate matrix}), we obtain the charge stability diagram shown in Fig.~\ref{fig1}b. We work in the regime where two holes are confined in the DQD.

Figure~\ref{fig1}c shows a simulated energy spectrum of two holes in a DQD as a function of detuning $\varepsilon$, defined along the axis indicated in Fig.~\ref{fig1}b (white dashed line), with $\varepsilon=0$ at the $(2,0)$--$(1,1)$ charge transition. For large positive or negative $\varepsilon$, both holes occupy the ground singlet states $S(0,2)$ or $S(2,0)$ of the same QD. In the intermediate regime, where each QD hosts a single hole, and for $J>\Delta{E}_\mathrm{Z}$, the low-energy states consist of the singlet $S(1,1)$ and triplets $T_0(1,1)$, $T_-(1,1)$ and $T_+(1,1)$. The logical qubit is encoded in the $\{\ket{S},\ket{T_0}\}$ subspace, while the polarized triplets are separated from this manifold by $\bar{E}_\mathrm{Z}=\frac{1}{2}(g_1+g_2)\mu_B B$, with $g_1$, $g_2$ the effective hole $g$-factors. In the regime where $\Delta E_\mathrm{Z} > J$, the eigenstates transition from $\{\ket{S},\ket{T_0}\}$ to the product spin states $\ket{\uparrow\downarrow}$ and $\ket{\downarrow\uparrow}$.

To probe this spectrum experimentally, we initialize $\ket{S(2,0)}$ deep in the $(2,0)$ region and apply a voltage pulse of variable amplitude and duration along the detuning axis. As the pulse traverses the relevant anticrossings, the singlet state undergoes coherent diabatic evolution, generating oscillations shown in Fig.~\ref{fig1}d that map the level structure. Readout is performed via (latched) Pauli spin blockade, where singlet and triplet states map onto distinct charge configurations~\cite{Kelly2025,HarveyCollard2018} (Extended Data Fig. 1). Near the center of the diagram, the $S$--$T_0$ states hybridize with a coupling strength of $\Delta E_\mathrm{Z} =\Delta g\mu_B B$, where $\Delta g=g_1-g_2$. On either side of the $S$--$T_0$ anticrossing, additional oscillations are observed that originate from the $S(2,0)$--$T_-$ and $S(0,2)$--$T_-$ anticrossings~\cite{Jirovec2021}.

In Fig.~\ref{fig1}e, we activate the exchange energy $J$ as shown in an exchange-`fingerprint' measurement~\cite{Reed2016}. We apply simultaneous square pulses to the detuning and virtual interdot barrier gate $V_\mathrm{B12}$ then measure the singlet probability. $V_\mathrm{B12}$ tunes the interdot tunnel coupling and hence $J$, shifting the $S(2,0)$--$T_-$ and $S(0,2)$--$T_-$ anticrossing toward the middle detuning point, near the center of the $(1,1)$ region~\cite{Jirovec2022,Zhang2024}. In Fig.~\ref{fig1}d-e, the observed oscillation between $S$ and $T_0$ states is symmetric around this middle detuning point, $\varepsilon \approx 16$\,mV. We set this as the manipulation point for all DST operations to reduce sensitivity to detuning fluctuations~\cite{Reed2016}.

To further characterize the spin states at the manipulation point, we perform gate-driven resonant spectroscopy. The spins are first initialized into the $\ket{\downarrow\downarrow}$ state (also $T_-$) through an adiabatic transfer from the singlet ground state $S(2,0)$~\cite{Kelly2025,Tsoukalas2026}. An RF drive of variable frequency and duration is then applied to the interdot barrier $V_\mathrm{B12}$, followed by a symmetric readout sequence. In Fig.~\ref{fig1}f, we observe coherent rotations of each individual spin (upper panel: $\ket{\downarrow\downarrow}\leftrightarrow\ket{\uparrow\downarrow}$, and lower panel: $\ket{\downarrow\downarrow}\leftrightarrow\ket{\downarrow\uparrow}$) induced by the RF voltages, i.e. electric-dipole spin resonance (EDSR) \cite{Hendrickx2020}. Throughout this work, resonant driving by RF is used solely for spectroscopy to map out the relevant energy levels, while all coherent qubit operations are implemented using baseband voltage pulses.

\begin{figure*}[tbp]
\centering
\includegraphics[width=\textwidth]{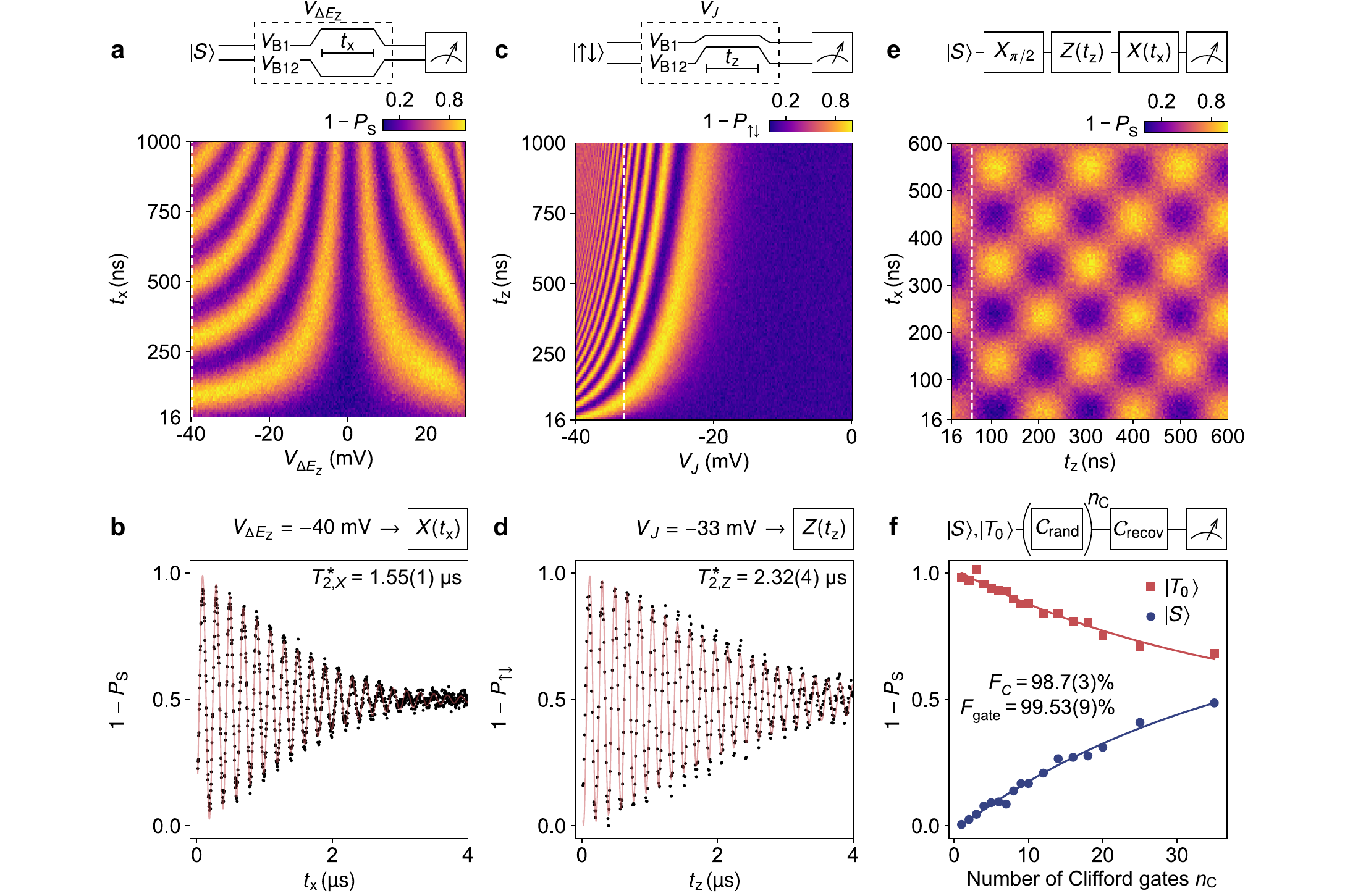}
\caption{
\textbf{Orthogonal two-axis control of the DST qubit.}
\textbf{a}, Coherent $X$ rotations generated by pulsing the virtual gate $V_{\Delta E_\mathrm{Z}}$, which consists of equal and opposite voltage pulses applied to the virtual gates $V_\mathrm{B1}$ and $V_\mathrm{B12}$, as illustrated in the pulse sequence shown above.
\textbf{b}, $X$ gate oscillations measured at $V_{\Delta E_\mathrm{Z}}=-40$\,mV (white dashed line in \textbf{a}). An exponential fit (red solid line) yields a coherence time of $T^*_{2,X}=1.55(1)$\,\textmu s.
\textbf{c}, Coherent $Z$ rotations generated by pulsing the virtual gate $V_{J}$, consisting of simultaneous voltage pulses applied to $V_\mathrm{B1}$ and $V_\mathrm{B12}$ with a ratio of 0.5 to 1, as illustrated in the pulse sequence shown above.
\textbf{d}, $Z$ gate oscillations measured at $V_{J}=-33$\,mV (white dashed line in \textbf{c}). An exponential fit (red solid line) yields a coherence time of $T^*_{2,Z}=2.32(4)$\,\textmu s.
\textbf{e}, Demonstration of two-axis control through the sequential application of a $Z$ gate of duration $t_\mathrm{z}$ and an $X$ gate of duration $t_\mathrm{x}$, as illustrated in the pulse sequence shown above. An $X_{\pi/2}$ pulse initializes the qubit along the $Y$-axis of the Bloch sphere. The $X$ gate oscillations vanish after a $Z_{\pi/2}$ rotation ($t_z= 48$ ns,  white dashed line), indicating alignment of the qubit state with the $X$-axis.
\textbf{f}, Randomized benchmarking of single-qubit gates in the singlet--triplet subspace. Average physical gate and Clifford gate fidelities of $99.53(9)\%$ and $98.7(3)\%$ are extracted.
}
\label{fig3}
\end{figure*}

\subsection{State space of a degenerate ST\texorpdfstring{$_0$}{0} qubit}

In the $\{\ket{S},\ket{T_0} \}$ basis, the ST qubit Hamiltonian can be written as
\begin{equation}
H = -\frac{J}{2}\sigma_z + \frac{\Delta E_\mathrm{Z}}{2}\sigma_x.
\label{eq:st0_hamiltonian}
\end{equation}

The Hamiltonian in Eq.~\ref{eq:st0_hamiltonian} is visually represented by the Bloch sphere in Fig.~\ref{fig2}a. The logical states $\ket{S}$ and $\ket{T_0}$ define the $Z$ axis and are split by the exchange energy $J$, whereas the antiparallel spin states $\ket{\uparrow\downarrow}$ and $\ket{\downarrow\uparrow}$ define the orthogonal $X$-axis and are separated by the difference in Zeeman energies $\Delta E_\mathrm{Z}$. Consequently, a degenerate idle qubit is obtained when $J$ and $\Delta E_\mathrm{Z}$ simultaneously vanish.

Hole spin qubits in Ge/SiGe have a highly anisotropic $g$-tensor, with the $g$-factor along the growth direction (out-of-plane) approximately two orders of magnitude larger than the in-plane one~\cite{Hendrickx2024}. Figure~\ref{fig2}b shows the resonance frequencies of the two spins as a function of the out-of-plane magnetic field angle $\theta$, measured using resonant spectroscopy at a fixed magnetic field amplitude $|B|=40$\,mT. As the field angle is varied from in-plane to out-of-plane, the spin resonance frequencies follow the Zeeman energies $\propto\sqrt{(g_\parallel B_\parallel)^2 + (g_\perp B_\perp)^2}$, tracing out two parabola-like curves, one for each spin. The frequency minima of the parabolas correspond to the in-plane $g$-tensor orientations of the two hole spins, which are slightly misaligned~\cite{Dijkema2026}. We define $\theta=0^\circ$ at the midpoint between the two $g$-factor minima as the in-plane direction of the two-spin system. At $\theta=0.8^\circ$ the two parabolas intersect, indicating that the two spin $g$-factors are equal, i.e. $\Delta E_\mathrm{Z}=0$. This magnetic field orientation is therefore chosen as the operating point of the DST qubit.

We next investigate the electrical tunability of $\Delta E_\mathrm{Z}$ with gate voltages. Figure~\ref{fig2}c shows resonant spectroscopy measurements acquired as a function of the virtual barrier voltage $V_\mathrm{B1}$. The two spin resonances, $\ket{\downarrow\downarrow}$ to $\ket{\downarrow\uparrow}$ and $\ket{\downarrow\downarrow}$ to $\ket{\uparrow\downarrow}$, disperse linearly with $V_\mathrm{B1}$ but with different slopes: $V_\mathrm{B1}$ predominantly affects the resonance frequency of the spin under $\mathrm{P}_1$ while leaving that under $\mathrm{P}_2$ comparatively unchanged. 
As a result, $\Delta E_\mathrm{Z}$ can be tuned at a rate of approximately $0.1\,\mathrm{MHz\,mV^{-1}}$.
Importantly, the transition frequency of each spin remains independent of the state of the neighboring spin, indicating that tuning $V_\mathrm{B1}$ does not generate a measurable exchange interaction (Extended Data Fig.~2).
Electrical control of $\Delta E_\mathrm{Z}$ will be an important capability for compensating local variability in a large-scale device~\cite{carballido2025compromise,bassi2026optimal}. We return to a detailed discussion on the $g$-tensor tunability in Section~\ref{sec:Tuning_the_degenerate_point}.

Similar to $\Delta E_\mathrm{Z}$, the exchange interaction $J$ can be tuned by gate voltage and characterized by resonant spectroscopy measurements, this time as a function of the middle barrier voltage $V_\mathrm{B12}$ in Fig.~\ref{fig2}d. In contrast to $\Delta E_\mathrm{Z}$, the spectrum displays a pronounced non-linearity associated with a $J$-dominated Hamiltonian in the regime where $J \gg \Delta E_\mathrm{Z}$. The frequency branch corresponding to the singlet-like state shifts exponentially with barrier voltage, whereas the triplet-like branch remains almost unaffected. From the difference of the two branches, we extract an exchange energy $J$ ranging from $0$ MHz up to $30\,\mathrm{MHz}$ for the barrier values $V_\mathrm{\text{B12}}$ = 25\,mV to $-20$\,mV. For $V_\mathrm{\text{B12}}$ > 25\,mV, the two frequency branches again split, as highlighted in the inset of Fig.~\ref{fig2}d. Since there is no $J$ in this regime due to exponentially suppressed tunneling, this splitting is attributed to $V_\mathrm{B12}$ also influencing $\Delta E_\mathrm{Z}$, albeit less strongly than $V_\mathrm{B1}$. Combining both $V_\mathrm{B12}$ and $V_\mathrm{B1}$ gives full electrical control over the parameters $\Delta E_\mathrm{Z}$ and $J$, which can be turned off and on by demand.

\subsection{Orthogonal control of the DST qubit}

In a degenerate qubit, the computational basis is defined by the initialization and readout protocol rather than by an intrinsic energy splitting. In our case, both the $\ket{S(1,1)}$ and $\ket{\uparrow\downarrow}$ states can be prepared using baseband voltage pulses alone, following the procedure described in Ref.~\cite{CovaFarina2025} and discussed in detail in Methods~\ref{Methods:initialization} using varying ramp rates and Hamiltonian adiabatic evolution. Furthermore, we define two effective control voltages, $V_J$ and $V_{\Delta E_\mathrm{Z}}$, by linear combinations of $V_\mathrm{B1}$ and $V_\mathrm{B12}$ to access $J$ and $\Delta E_\mathrm{Z}$ independently (Methods~\ref{Methods:initialization}). This virtualization method allows for the orthogonal control of the DST qubit.

We first consider control of $\Delta E_\mathrm{Z}$, which generates rotations around the $X$-axis of the DST Bloch sphere of Fig.~\ref{fig2}a. In the low-tunnel-coupling regime ($J =0$), the side barrier gates modify the individual spin frequencies without generating exchange interaction (Fig.~\ref{fig2}c-d). Fig.~\ref{fig3}a shows coherent $X$ rotations of an initialized $\ket{S}$ state generated by pulses on $V_{\Delta E_\mathrm{Z}}= (-1\cdot V_\mathrm{B12}, 1\cdot V_\mathrm{B1})$. The oscillation frequency increases linearly with increasing pulse amplitude, consistent with the Zeeman energy difference observed in spectroscopy (Fig.~\ref{fig2}c, Extended Data Fig.~2). The asymmetry of the map arises from a contribution of $V_\mathrm{B12}$ in $V_{\Delta E_\mathrm{Z}}$ to extend the accessible range of $\Delta E_\mathrm{Z}$ without requiring large pulse amplitudes. This contribution leads to a finite $J$ for positive $\Delta E_\mathrm{Z}$ (Extended Data Fig.~2 and 3). By setting $V_{\Delta E_\mathrm{Z}} = -40\,\mathrm{mV}$, we achieve an $X_{\pi}$ gate of 100\,ns, as can be seen from the $X$ Rabi oscillations shown in the linecut in Fig.~\ref{fig3}b.

We next implement rotations about the $Z$-axis by activating the exchange interaction. Because pulses applied to the middle barrier also induce a small change in $\Delta E_\mathrm{Z}$ (Fig.~\ref{fig2}d inset), a compensation pulse must be simultaneously applied to $V_\mathrm{B1}$ to cancel it. From measurements performed in the regime where $J$ and $\Delta E_\mathrm{Z}$ are comparable, we extract a compensation ratio of approximately $\Delta E_\mathrm{Z}^{B12}/\Delta E_\mathrm{Z}^{B1} \simeq0.5$ (Extended Data Fig.~4), consistent with the value obtained in the $J=0$ limit (Fig.~\ref{fig2}d). Figure~\ref{fig3}c shows coherent $J$ rotations between the $\ket{\downarrow\uparrow}$ and $\ket{\uparrow\downarrow}$ states generated using the virtual gate $V_J = (1\cdot V_\mathrm{B12}, 0.5\cdot V_\mathrm{B1})$. The oscillations maintain high contrast across the full tuning range, indicating that the rotation axis remains well aligned with the exchange direction (Extended Data Fig.~4). Setting $V_J =- 33\,\mathrm{mV}$ results in a $Z_{\pi}$ gate of $\sim100$\,ns, as can be seen from the $Z$ Rabi oscillations shown in the linecut in Fig.~\ref{fig3}d.

In Fig.~\ref{fig3}e, we perform a sequence to demonstrate arbitrary rotations in the ST Bloch sphere~\cite{Tsoukalas2026} . Starting from $\ket{S}$ and applying an $X_{\pi/2}$ pulse, we place the system's state pointing along the $Y$-axis of the Bloch sphere. Subsequently, we apply a $Z$ pulse for duration $t_z$ and then an $X$ pulse for $t_x$, before reading out the $\ket{S}$ state population.
This sequence effectively prepares and reads out different states on the Bloch sphere. For $t_z = t_{\pi/2}\sim48$\,ns (dashed line in Fig.~\ref{fig3}e), the qubit state points along the $X$-axis, and therefore no oscillations are observed against $t_x$, with the state probability remaining at approximately $50\%$. Finally, the independent control of $Z$ and $X$ rotation axes enables arbitrary combinations of the two; one example is the Hadamard gate, obtained by setting $\Delta E_\mathrm{Z} = J$, as shown in Extended Data Fig. 5.

\begin{figure*}[tbp]
\centering
\includegraphics[width=\linewidth]{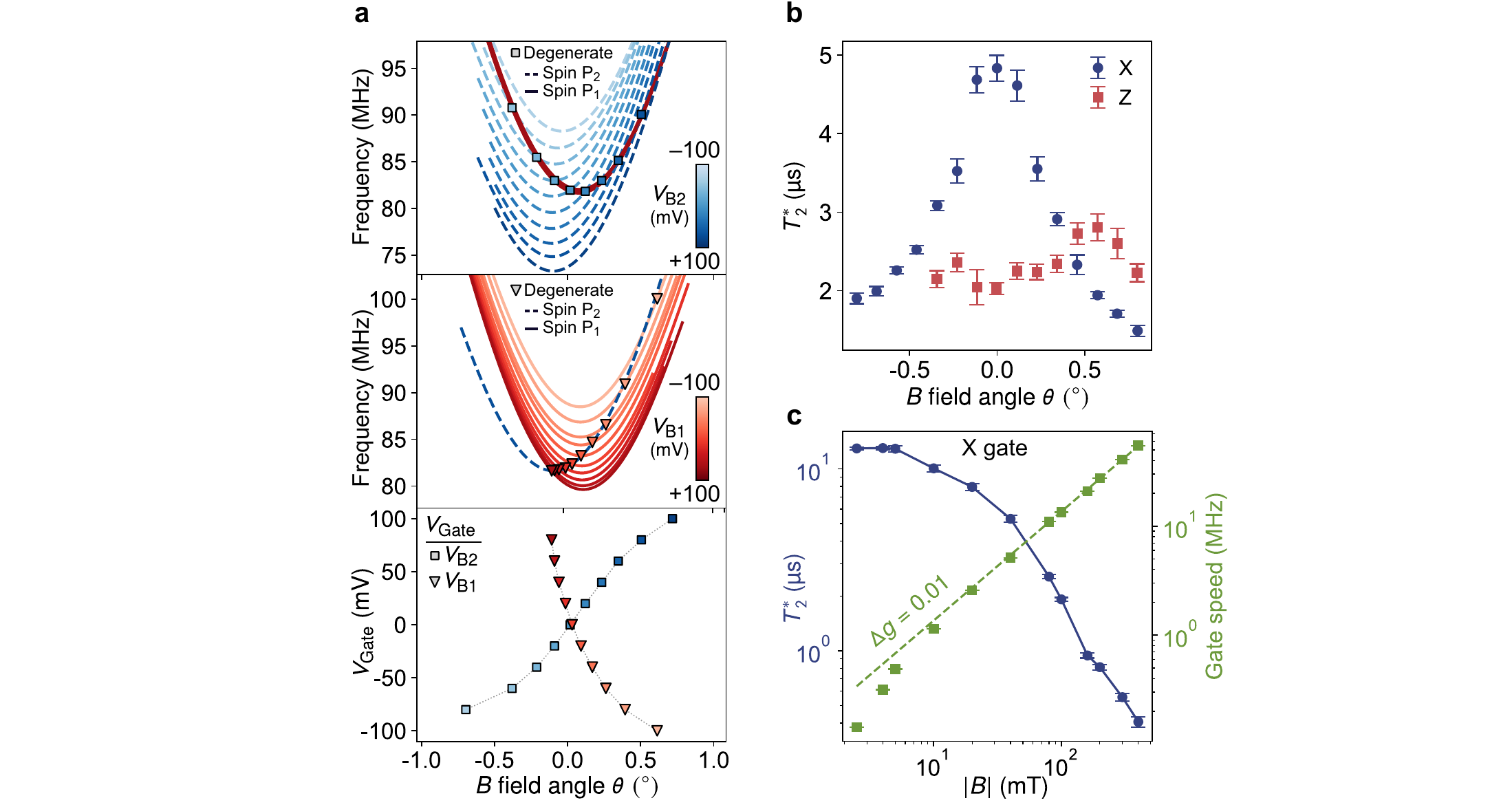}
\caption{
\textbf{Magnetic-field dependence of the DST coherence.}
\textbf{a}, Extracted spin-resonance frequencies of the two holes as a function of magnetic field angle for different values of $V_\mathrm{B2}$ (top) and $V_\mathrm{B1}$ (middle). The degenerate points, where the two spin frequencies coincide, are indicated by square and triangular markers, respectively. The resulting dependence of the degenerate point on gate voltage is summarized in the bottom panel. For visualization purposes, the less tunable $g$-tensor has been renormalized to its average value in the cumulative plots.
\textbf{b}, Coherence times $T^*_{2,X}$ (blue markers) and $T^*_{2,Z}$ (red markers) measured at the different degenerate points corresponding to the magnetic field orientations shown in \textbf{a}. Only values extracted for well-calibrated gates are plotted.
\textbf{c}, At the in-plane angle $\theta=0^\circ$, $X$-rotation coherence time $T^*_{2,X}$ (blue markers, left y-axis) and corresponding oscillation frequency obtained at fixed $V_{\Delta E_\mathrm{Z}}$ (green markers, right y-axis) as a function of magnetic field magnitude. The dashed green line corresponds to $\Delta E_\mathrm{Z}=\Delta g\,\mu_\mathrm{B}|B|$ with $\Delta g\approx 0.01$.
}
\label{fig4}
\end{figure*}

\subsection{Metrics of the DST qubit}

For the DST qubit, there are two relevant coherence times associated with the distinct Hamiltonians that generate the gate operations, namely $T^*_{2,X}$ and $T^*_{2,Z}$ corresponding to $\Delta E_\mathrm{Z}$ and $J$ rotations. The equivalent of the free-induction decay time, $T_2^*$, is the coherence time at the degenerate point where both $\Delta E_\mathrm{Z}$ and $J$ are zero. At this point, $J$ is exponentially suppressed for small voltage fluctuations and the dominant noise affecting the coherence of the DST is in $\Delta E_\mathrm{Z}$. We find that $T^*_{2,X}=T_2^*$ for all $X$-rotation frequencies in Fig.~\ref{fig3}a, indicating that the noise of $\Delta E_\mathrm{Z}$ is independent of the $\Delta g$ value (Extended Data Fig. 6a). By fitting decaying exponentials to the $X$- and $Z$-axis Rabi oscillations shown in Fig.~\ref{fig3}b and Fig.~\ref{fig3}d, respectively, we extract coherence times of $T_2^*=T^*_{2,X}=1.55$\,\textmu s and $T^*_{2,Z}=2.32$\,\textmu s. The dominant $X$-axis noise at the degenerate point is further revealed by a Hahn echo sequence, in which the application of a $Z_\pi$ gate at the midpoint of the idle interval extends the coherence time to $T_2^{\mathrm{H}} = 4$\,\textmu s  while an $X_\pi$ echo shows no improvement (Extended Data Fig. 6b). Further improvement of the coherence time ($T_2^{\mathrm{CPMG}}$) up to 40\,\textmu s can be achieved with dynamical decoupling sequences like Carr–Purcell–Meiboom–Gill (Extended Data Fig. 6c).

The fidelity of the DST qubit single-qubit gates is estimated through randomized benchmarking (RB)~\cite{Zhang2024,Tsoukalas2026} using the Clifford gate set based on $\{X_\pi, X_{\pi/2}, Z_\pi, Z_{\pi/2}, I \}$ (Methods~\ref{Methods:RB}). The obtained RB data are shown in Fig.~\ref{fig3}f. We extract an average physical gate fidelity of $F_{\text{gate}} = 99.53(9)\%$ and an average Clifford gate fidelity of $F_\mathrm{C} = 98.7(3)\%$. These fidelities are on par with those achieved in state-of-the-art singlet--triplet qubits~\cite{Zhang2024, Tsoukalas2026,takeda2020resonantly,cerfontaine2020closed}.

\subsection{Tuning the degenerate point in the magnetic field sweet spot}
\label{sec:Tuning_the_degenerate_point}
For further insight, we systematically vary the virtual barrier voltages $V_\mathrm{B1}$ and $V_\mathrm{B2}$  and measure their effects on the $g$-factors by resonant spectroscopy as in Fig.~\ref{fig2}b. In Fig.~\ref{fig4}a(top, middle), we show cumulative results acquired for different values of $V_\mathrm{B1}$ and $V_\mathrm{B2}$ ranging from -100 mV to 100 mV (Extended Data Figs. 7-8). A clear trend emerges where each barrier gate predominantly tunes the in-plane $g$-factor magnitude (vertical shift) of its nearest spin without significantly changing the in-plane orientation (horizontal shift). This behavior is consistent across the full $V_\mathrm{B2}$ range, whereas for more negative $V_\mathrm{B1}$ values, the influence on both spins becomes comparable, saturating its ability to shift the degenerate point. The observed $g$-factor tunability is attributed to light-hole–heavy-hole mixing, in which electrostatic confinement (i.e., confinement potential shape)~\cite{Hendrickx2024,Sommer2026, Bosco2021} and inhomogeneous strain~\cite{Scappucci2021,liles2021electrical,AbadilloUriel2023,Seidler2025} due to heterostructure growth play a significant role. 

The degenerate operating point, marked by the intersection of the two sets of parabolas (square/triangle symbols), is tuned by $V_\mathrm{B1}$ and $V_\mathrm{B2}$ across a wide range of field angle $\theta$ shown in Fig.~\ref{fig4}a(bottom). The operating magnetic field, in both orientation and magnitude, strongly affects the performance of germanium hole qubits~\cite{Hendrickx2024}. For the DST qubit, we first vary the field angle at a fixed magnitude $|B|= 40$\,mT. For each field angle $\theta$, we retune to the degenerate condition and then perform both X and Z rotations with similar Rabi frequencies ($\sim5$\,MHz). The extracted coherence times for both control axes are shown in Fig.~\ref{fig4}b. $T^*_{2,Z}$ shows no clear
dependence on $\theta$, remaining at $2$--$2.5$\,\textmu s. In contrast,
$T^*_{2,X}$
depends strongly on field direction, peaking at approximately
$5$\,\textmu s for $\theta = 0^\circ$ (in-plane) and decreasing toward the
value measured in Fig.~\ref{fig3}b as the degenerate point moves further
out of plane.  These distinct trends reveal the different noise sources
limiting each gate. The $Z$ gate, for which $J$ is the dominant energy
scale, is mainly limited by charge (tunneling) noise, consistent with
Fig.~\ref{fig3}c: faster $Z$ rotations decohere more rapidly because a
more negative $V_J$ increases $J$ and thus the sensitivity to charge noise
($\partial J/\partial V_J \propto J$). By contrast, the X gate, driven by $\Delta E_\mathrm{Z}$ is limited by magnetic noise through the hyperfine interaction, which is weaker when $B$ points in-plane~\cite{bosco2021fully, Hendrickx2024}. Charge noise also contributes, since $\Delta E_\mathrm{Z}$ is
voltage-controlled, but its effect is independent of
$V_{\Delta E_\mathrm{Z}}$ value because $\partial\Delta E_\mathrm{Z}/\partial V_{\Delta E_\mathrm{Z}} \approx
\mathrm{constant}$.

Next, we fix $\theta=0^\circ$ and vary $\abs{B}$ from $400\,\mathrm{mT}$ down to $2\,\mathrm{mT}$ and extract $T^*_{2,X}$. As shown in Fig.~\ref{fig4}c, we observe a rapid increase in $T^*_{2,X}$ as the magnetic field is reduced, saturating at $12.5$\,\textmu s for $\abs{B}$ below $5\,\mathrm{mT}$. This behavior is consistent with previous studies of Loss-DiVincenzo hole spin qubits performed at similar magnetic field amplitudes~\cite{Wang2024,Hendrickx2024}. In the same figure, we also plot the $X$ gate speed at a fixed $V_{\Delta E_\mathrm{Z}}=-40$\,mV, scaling linearly with the magnetic field magnitude. The green dashed line marks the fit for an effective value $\Delta g \approx 0.01$ induced by the gate voltage. Despite the improved coherence times at low fields, no clear advantage in gate fidelity or gate quality factor is obtained for magnetic fields below approximately $40\,\mathrm{mT}$. An additional measurement under a lower field amplitude $|B|$ = 24\,mT at in-plane angle $\theta=0^\circ$ is provided in Extended Data Fig. 9 with similar behavior to that in Fig.~\ref{fig3}.

\section{Discussion}
In summary, we demonstrated orthogonal two-axis control and a degenerate idle point of a singlet–triplet qubit formed by two hole spins in germanium using only baseband electrical pulses. An average physical single-qubit gate fidelity of 99.53\% was achieved (gate set: $\{X_\pi, X_{\pi/2}, Z_\pi, Z_{\pi/2}, I \}$), while two distinct coherence times were identified for the two orthogonal rotation axes controlled by $J$ and $\Delta E_\mathrm{Z}$. By studying how voltages applied to neighboring gate electrodes modify the $g$-factors of each hole spin, we were able to efficiently tune the degenerate point across a range of magnetic field orientations. This tunability enables operation at an in-plane angle where the reduced noise sensitivity leads to a significant increase of the coherence time.

Looking ahead, the DST qubit has promising characteristics for scalability and would benefit strongly from continued advances in material development and device optimization. For scaling to larger arrays, the ability to electrically tune the degenerate point over a wide range of magnetic field orientations suggests that multiple DST qubits can be operated simultaneously, while dedicated $g$-factor tuning gates would further simplify their operation. In addition, continued improvements in material quality~\cite{Lodari2022,Massai2024,Stehouwer2025} can lead to suppressed hyperfine noise through isotopic purification of germanium~\cite{Itoh1993} and improved $g$-factor uniformity~\cite{Stehouwer2025}, thereby further relaxing tuning requirements and enhancing device performance.

However, perfect $g$-tensor uniformity across all QDs may not be the optimal goal, as engineering local $g$-tensor variations could provide an additional tuning knob for implementing high-fidelity two-qubit operations. Specifically, during exchange-based gates between two ST qubits (ST$_1$--ST$_2$), the leakage probability to states outside the computational subspace scales as $P_{\mathrm{leak}} \sim (J/\Delta E_\mathrm{Z}^{\mathrm{ST12}})^2$. Therefore, engineering significantly different average Zeeman energies, $E_\mathrm{Z}^{\mathrm{ST1}}$ and $E_\mathrm{Z}^{\mathrm{ST2}}$, for neighboring ST qubits suppresses this leakage~\cite{Spethmann2024,Ni2025}, as discussed in detail in Methods~\ref{Methods:Two-qubit gates}. The demonstrated electrical tunability of the $g$-factor offers one possible route to achieving this.
Additional approaches could include strain engineering~\cite{CorleyWiciak2023} and device geometry, for example, alternating dot sizes~\cite{Bosco2021,valvo2025electrically} or implementing devices with multiple quantum wells~\cite{tidjani2023vertical,tidjani2025three} with different average $g$-factors.

\section*{Methods}
\renewcommand{\thesection}{Sec.}
\renewcommand{\thesubsection}{\Alph{subsection}}
\setcounter{subsection}{0}

\subsection{Material growth and device fabrication}
\label{Methods:fabrication}
The device is fabricated similarly to the one described in Ref.~\cite{John2025}, but on a silicon wafer instead of a germanium wafer. It is built on a Ge/SiGe heterostructure~\cite{sammak2019shallow} containing a 16 nm thick germanium quantum well, located 55 nm beneath the semiconductor-oxide interface, grown by chemical vapor deposition on a silicon substrate. Fabrication begins with the formation of ohmic contacts via platinum deposition followed by diffusion.
The gate stack is then built up in three stages, with each layer made of palladium deposited at room temperature: first the 20 nm barrier gate layer, then the 30 nm screening gate layer, and finally the 40 nm plunger gate layer. A 7 nm aluminum oxide layer separates the barrier gates from the underlying heterostructure, and an additional 5 nm oxide layer is applied after both the barrier and screening gate depositions.

\subsection{Experimental setup}
\label{Methods:setup}
 
The device is loaded into a dilution refrigerator (Bluefors LD400) with a base temperature of 12\,mK and a 9\,T--1\,T--1\,T superconducting vector magnet (American Magnetics). The setup has 96 DC lines and 42 RF lines. The DC lines are filtered at the mixing chamber (MXC) stage with silver-powder filters (Basel Precision Instruments) and custom LC and RC filters with a lowest cutoff frequency of around 1\,kHz. For the germanium device, we measure an effective electron temperature around 120\,mK by standard Coulomb-blockade thermometry on one of the sensor dots (Extended Data Fig. 10). The DC voltages are controlled by custom-made digital-to-analog converter (DAC) boxes. Electronic transport is measured using a lock-in amplifier (Stanford Research SR830) and a current preamplifier (Basel Instruments SP983c). The RF lines are made from low-loss CuNi coaxial cables, thermally anchored at each fridge stage (50\,K, 4\,K, 1\,K, MXC), with total attenuation ranging from $-30$\,dB to $-60$\,dB. The RF voltages are controlled by a Quantum Machines OPX1000 LF FEM. A minimum ramp duration of 16\,ns is introduced to all baseband pulses. The AC and DC lines are combined by bias tees on the printed circuit board (PCB). The input and output lines for RF reflectometry are isolated by a directional coupler (Mini-Circuits ZEDC-15-2B). The reflected RF signal is amplified with 30\,dB of gain by a cryogenic high-electron-mobility transistor (HEMT) amplifier (Low Noise Factory LNC0.2\_3B) at the 4\,K stage, and by a further 30\,dB at room temperature.
 
\subsection{RF reflectometry}
\label{Methods:reflectometry}
We perform charge and spin-state readout by RF reflectometry using a nearby charge sensor (Fig.~\ref{fig1}a). On the PCB, the sensor dot is connected to an LC tank circuit with a resonance frequency of 276\,MHz. The sensor is tuned into the few-hole regime using its plunger and barrier gates. In operation, it is parked on the flank of a Coulomb peak, where its resistance depends sensitively on the nearby charge environment. We monitor the change in the reflection coefficient at the LC resonance as a proxy for the change in sensor resistance. During qubit operation, different spin states are mapped to different charge states---$(1,1)$ or $(2,0)$---by Pauli spin blockade, producing contrast in the RF reflection signal (Extended Data Fig. 1). We typically use a readout time of 3\,\textmu s and average many shots to obtain clear contrast between spin states. We perform calibration to assign the averaged readout signal to the corresponding spin-state probability, as shown in the histogram in Extended Data Fig. 1d. For each spin state, 100{,}000 single-shot measurements were taken, yielding two Gaussian-shaped histograms for the $S$ and $T_0$ states, respectively. The peak of each Gaussian agrees with its many-shot average value and can therefore be assigned a probability of 1 for that spin state. An averaged value falling between the two peaks is then normalized to a probability between 0 and 1.
 
\subsection{Virtual gate matrix}
\label{Methods:Virtual gate matrix}

The capacitive coupling between different gates is captured by the virtual gate matrix, which we construct in several layers. First, all the physical gates are virtualized against the sensor to maintain readout sensitivity at all times. Second, the detuning and chemical-potential gates, $v\varepsilon$ and $vU$, are defined as linear combinations of $v\text{P1}$ and $v\text{P2}$ that control, respectively, only the chemical-potential difference $(\mu_1 - \mu_2)$ and the average chemical potential $(\mu_1 + \mu_2)/2$, where $\mu_1$ and $\mu_2$ are the two dot chemical potentials. We always operate the qubit in the $(1,1)$ region at the middle detuning, $\varepsilon = 16$\,mV, where it is insensitive to detuning noise to first order ($\mathrm{d}J/\mathrm{d}\varepsilon = 0$) \cite{Reed2016}. Third, all other barrier gates---$B_{1}$, $B_{12}$, and $B_2$---are virtualized against $v\varepsilon$ and $vU$, which are held constant during qubit gate operation. Finally, a linear combination of $vB_1$, $vB_{12}$, and $vB_2$ provides orthogonal control of the X and Z gates, denoted $V_{\Delta E_\mathrm{Z}}$ and $V_J$. The full gate matrix, including all layers, is shown in Supplementary Table 1. All gate variables in all figures are virtual gates; the $v$-prefix is omitted for simplicity unless noted otherwise.

\subsection{Initialization, readout and gate orthogonality}
\label{Methods:initialization}

The spin state of the DST qubit can be initialized in either the $S$--$T_0$ basis or the $\uparrow\downarrow/\downarrow\uparrow$ basis, using different ramp speeds and initialization sequences. In both cases, we start in the charge $(2,0)$ region near the $(1,0)$ boundary, where the $T_1$ spin relaxation rate is fast,  and then move adiabatically to the middle of the $(2,0)$ region. The dwell time ($>10$\,\textmu s) is long enough to ensure that the two spins relax to their true ground state, $S(2,0)$.
 
For singlet initialization, we apply a fast ramp (16\,ns) from the middle of $(2,0)$ into the middle of the $(1,1)$ region, while simultaneously ramping the exchange $J$ and $\Delta E_\mathrm{Z}$ down to zero. Because the ramp rate $\gg \Delta_{ST_-}$ (singlet--$T_-$ coupling) and $\Delta E_\mathrm{Z}$, this produces a singlet initialization in the degenerate regime ($J = 0$, $\Delta E_\mathrm{Z} = 0$). Readout is performed by reversing the ramp steps at the same speed and then pulsing to the Pauli spin blockade (PSB) window. In Extended Data Fig. 3b-f, we confirm this procedure by applying an X gate ($\Delta E_\mathrm{Z}$) and a Z gate ($J$ gate) to the singlet state: strong oscillations in the $X$ gate map indicate proper singlet--triplet rotation, while no contrast appears in the $Z$ gate map because the singlet is an eigenstate of the $J$ gate.
 
For $\uparrow\downarrow$ initialization, we again apply a fast ramp (16\,ns) from the middle of $(2,0)$ into the middle of the $(1,1)$ region, ramping $\Delta E_\mathrm{Z}$ down to zero but keeping the exchange $J$ on. As before, this produces a singlet initialization, but now with $J > 0$, since the ramp rate $\gg \Delta_{ST_-}$ and $\Delta E_\mathrm{Z}$. We then adiabatically ramp (6\,\textmu s) $J$ down to zero while ramping $\Delta E_\mathrm{Z}$ up; the state therefore evolves into the true eigenstate of the Hamiltonian and becomes $\uparrow\downarrow$ ($J = 0$, $\Delta E_\mathrm{Z} > 0$). Finally, we turn off $\Delta E_\mathrm{Z}$, yielding an $\uparrow\downarrow$ initialization in the degenerate regime ($J = 0$, $\Delta E_\mathrm{Z} = 0$). Readout reverses the initialization steps and then proceeds to the PSB location. As with the singlet, we confirm the $\uparrow\downarrow$ state by applying the same X gate ($\Delta E_\mathrm{Z}$) and Z gate ($J$ gate) in Extended Data Fig. 3a-b: now strong oscillations appear in the $Z$ gate map, indicating proper $\uparrow\downarrow/\downarrow\uparrow$ rotation, while no contrast appears in the $X$ gate map. Alternatively, the $\uparrow\downarrow$ state can be initialized by applying an $X_{\pi/2}$ gate after the singlet initialization, followed by $Z_{\pi/2}$ (Fig. 3e, dashed white line). This second method gives the same result as the first.
 
These two initialization/readout methods for the singlet and $\uparrow\downarrow$ states also confirm the orthogonality of our $X$ and $Z$ gates. Any X component in the applied Z gate would appear as oscillatory contrast in the $S/T_0$ initialization/readout, and likewise any Z component in the $X$ gate in the $\uparrow\downarrow/\downarrow\uparrow$ initialization/readout. Combining this baseband-contrast approach (Extended Data Fig. 3-4) with EDSR spectroscopy (Extended Data Fig. 2) and the arbitrary rotation checkerboard-style map (Fig.~\ref{fig3}e), we can confirm orthogonal operations at a desirable speed ($t_\pi \approx 100$\,ns): $V_{\Delta E_\mathrm{Z}} = (-1\cdot V_\mathrm{B12}, 1\cdot V_\mathrm{B1})$ and $V_J = (1\cdot V_\mathrm{B12}, 0.5 \cdot V_\mathrm{B1})$ shown in Fig.~\ref{fig3}.
 
\subsection{Randomized Benchmarking}
\label{Methods:RB}

We construct the 24 Clifford gates using a combination of $I$, $X_\pi$, $X_{\pi/2}$, $Z_\pi$, and $Z_{\pi/2}$ gates, as shown in Supplementary Table 2. On average, each Clifford gate requires 2.83 physical gates. The standard randomized benchmarking sequence proceeds as follows. We first initialize the state as either a singlet or a triplet, then apply $n_C$ Clifford gates chosen randomly from the 24 possibilities. A final Clifford gate is then applied to return the system to its original state before readout. For each value of $n_C$, we perform 200 such randomizations, and the averaged results for both singlet and triplet initialization are shown in Fig.~\ref{fig3}f. The singlet and triplet decays were fitted jointly to $P(n_C) = A p^{n_C} + B$, where $p$ is the shared depolarizing parameter (bounded between 0 and 1, with 1 corresponding to an ideal, error-free gate), and $A$ and $B$ are independent constants for each curve. We obtained $p = 0.973(5)$. The average Clifford gate fidelity is $F_\mathrm{C} = 1 - r_C = 1 - (1 - p)/2 = 98.7(3)\%$, and the average physical gate fidelity is $F_{\text{gate}} = 1 - r_C/2.83 = 99.53(9)\%$.

\subsection{Two-qubit gates and leakage channels}
\label{Methods:Two-qubit gates}

We consider two degenerate $S$--$T_0$ qubits with their corresponding single-spin basis
$\{\ket{\uparrow_1\downarrow_2}, \ket{\downarrow_1\uparrow_2}\}$ and
$\{\ket{\uparrow_3\downarrow_4}, \ket{\downarrow_3\uparrow_4}\}$.

A controlled-phase $CZ$ gate is implemented by activating the exchange interaction $J$ between spins 2 and 3. In addition to generating the intended conditional-phase evolution, this interaction couples the computational subspace to leakage states outside the encoded manifold. 
The leakage states that are close in energy are the non-computational states with $\left<S_z\right>=0$.

These states are detuned from the computational manifold by the Zeeman-energy difference between the two degenerate $S$--$T_0$ qubits,
\begin{equation}
\Delta E_\mathrm{Z}^{\mathrm{ST12}} = (g_{\mathrm{ST},1} - g_{\mathrm{ST},2}) \mu_B B,
\end{equation}
where $g_{\mathrm{ST},1} = g_1 = g_2$ and $g_{\mathrm{ST},2} = g_3 = g_4$ due to the Zeeman degeneracy within each qubit. We neglect additional leakage to higher energy states arising from spin--orbit interaction, which is strongly suppressed by the large energy gap at finite magnetic field.

Within each leakage manifold, the dynamics can be described as an effective two-level system with coherent coupling $J$ and detuning $\Delta E_\mathrm{Z}^{\mathrm{ST12}}$. The resulting leakage probability is

\begin{equation}
P_{\mathrm{leak}}(t)
=
\frac{1}{1+\left(\frac{\Delta E_\mathrm{Z}^{\mathrm{ST12}}}{J}\right)^2}
\sin^2\!\left[
\frac{J t}{2\hbar}
\sqrt{1+\left(\frac{\Delta E_\mathrm{Z}^{\mathrm{ST12}}}{J}\right)^2}
\right].
\end{equation}

This expression shows that leakage is suppressed as the ratio $\Delta E_\mathrm{Z}^{\mathrm{ST12}}/J$ increases, with a maximum leakage probability scaling as $P_{\mathrm{leak}}^{\mathrm{max}} \sim (J/\Delta E_\mathrm{Z}^{\mathrm{ST12}})^2$ for $\Delta E_\mathrm{Z}^{\mathrm{ST12}} \gg J$. When $\Delta E_\mathrm{Z}^{\mathrm{ST12}}/J$ cannot be made sufficiently large or to increase the two-qubit gate speed, which scales with $J$, leakage can alternatively be eliminated by synchronizing the leakage oscillations with the duration of the  $CZ$ gate. For an exchange-based CZ gate, the gate time is $t_{\mathrm{CZ}} = \pi\hbar/(4J)$. Requiring that the leakage amplitude vanishes at $t_{\mathrm{CZ}}$ yields the condition
\begin{equation}
\frac{1}{8}\sqrt{1+\left(\frac{\Delta E_\mathrm{Z}^{\mathrm{ST12}}}{J}\right)^2} = n,
\qquad n \in \mathbb{N},
\end{equation}
which is satisfied for
\begin{equation}
J = \frac{\Delta E_\mathrm{Z}^{\mathrm{ST12}}}{\sqrt{(8n)^2 - 1}}.
\end{equation}
At these operating points, transient leakage occurs during the gate, but the system returns exactly to the computational subspace at the end of the CZ operation. Increasing $\Delta E_\mathrm{Z}^{\mathrm{ST12}}$ therefore both suppresses leakage and allows faster gate operation, while residual spin--orbit-induced couplings may introduce small corrections beyond this simplified model.

\section*{Acknowledgments}
JHU acknowledges fruitful discussions with Sander L. de Snoo, and Alexander Ivlev and technical support from Nicholas R. Poniatowski, James McArthur and Steven Sansone. JS acknowledges support from Klaus Ensslin. KT acknowledges support from the Swiss National Science Foundation through the Postdoc.Mobility fellowship (Grant no: P500-2\_235507).
This research was supported by the Army Research Office under Award Number W911NF-23-1-0110.

\section*{Competing Interests}
G.S. and M.V. are founding advisors of Groove Quantum BV and declare equity interests.

\section*{Data Availability}
All data will be available in the Zenodo data repository.

\bibliographystyle{apsrev4-1}
\bibliography{references}

\end{document}